# Do open source software developers listen to their users?

Arif Raza and Luiz Fernando Capretz
Western University
Department of Electrical and Computer Engineering
London, Ontario, Canada - N6A5B9
{araza7, lcapretz}@uwo.ca

## Abstract

In application software, the satisfaction of target users makes the software more acceptable. Open Source Software (OSS) systems have neither the physical nor the commercial boundaries of proprietary software, thus users from all over the world can interact with them. This free access is advantageous, as increasing numbers of users are able to access OSS; there are more chances of improvement. This study examines the way users' feedback is handled by OSS developers. In our survey, we have also inquired whether OSS developers consult professional usability experts to improve their projects. According to the results, majority of OSS developers neither consider usability as their top priority nor do they consult usability experts.

## Introduction

The International Organization for Standardization and the International Electrotechnical Commission (ISO/IEC 9126-1) classifies software quality attributes into six categories: functionality, reliability, usability, efficiency, maintainability and portability [1]. The standard ISO/IEC 9126-1 states that usability is *"the capability of the software product to be understood, learned, used and attractive to the user, when used under specified conditions."*

OSS has influenced almost every dimension of the software development field, thus indicating its significant progress and evolution. The most successful examples of this influence include the GNU/Linux operating system, the Apache HTTP server, the Mozilla Firefox internet browser, and the MySQL database system. The aspect and measurement of quality assurance as well as the post-release management of OSS projects are some of the areas where closed source proprietary software is superior.

Although user-centered designs are gaining popularity within OSS community, many design scenarios still do not consider usability as one of their primary goals. OSS is having an increasing diversity of users, including those with technical and non-technical backgrounds as well as those from varying cultures, each with their own needs, expectations and demands. Even in the environment of closed proprietary software, usability is a complicated issue; however, in OSS, it is even more difficult, especially considering that the domain is relatively newer with developers working on a voluntarily basis.

This survey-based research has been carried out in order to understand the way OSS developers seek users' feedback and how do they meet the expectations of their target audiences. We have also inquired about the possible role of usability experts in OSS environment.

**Open Source Software and Usability**

Open source software refers to software that is equipped with licenses providing current and future users with the right to use, inspect, modify, and distribute modified or unmodified versions of the software to others [2].

Zhao et al. [3] consider OSS usability improvement an important matter that necessitates added exploration. They stress OSS community to improve quality and usability of their products. They test a set of hypotheses in a controlled environment to explore effects of different components of on effectiveness and efficiency of OSS usability improvement. Çetin and Göktürk [4] consider OSS a major platform for collaborative and cooperative software development. They also call for more usable system in OSS environment. Otte et al. [5] also underscore the high rate of user contribution, user inspection and peer reviews in OSS culture. Bodker et al. [6] however, observe that OSS developers need to have a thorough realization of user expectations.

Referring to the international standards for usability, Bevan [7] maintains that although software usability can be integrated with quality using these standards, it would not assist in usability improvement unless it is given a higher priority. Nichols and Twidale [8] observe that traditionally there have been fewer usability experts in the OSS world. Iivari et al. [9] also call for expert opinions as well as realistic user opinion at an earlier OSS design phase. Indicating OSS usability as a multidimensional problem area, Çetin and Göktürk [4] also identify that neither are they aware of user requirements, nor do OSS developers consider usability as a primary objective of their projects. Cetin et al. [10] identify users, customers and developers as the major sources of bug reporting in OSS. They emphasize using experts' opinions to improve the OSS usability. Lee et al. [11] also carry out an empirical study to measure success of OSS projects and realize importance of "*software quality and user satisfaction.*"

According to Raza and Capretz [12], the challenges OSS is facing today include obtaining an enhanced understanding of contributors' opinions, taking on new design approaches to improve usability, and enumerating usability metrics.

**Research Methodology**

In our survey, we posted five questions (Q-1 to Q-5) as shown in Table 1. We explained to our participants that they have been asked to fill in this survey because they have participated in OSS development in the past 5 years. Q-1 was to determine which quality attribute has top priority from OSS developers' perspective. Q-2 and Q-3 were related to user feedback, and Q-4 and Q-5 were about usability experts' opinion.

Open source software projects deal with different domains of applications. Accordingly, we sent personalized emails to OSS developers of different projects on sourceforge.net. The projects differed in size and ranged from small-scale to large-scale projects. Subsequently, we sent our questionnaires to OSS developers working on projects in the categories of education, scientific / engineering, database, games / entertainment, text editors, development, testing, communications, and multimedia, as shown in Figure 1.

We assured the participants that our survey did not require their identity and would not be recorded. We received responses of 72 OSS developers altogether.

*Reliability and Validity Analysis of the Measuring Instrument*

The reliability of a measurement and the validity (the strength of the inference between the true value and the value of a measurement) are the two integral features of an empirical study. The reliability of the measurement scales is evaluated by using internal-consistency analysis, which is performed using the coefficient alpha [13]. In our analysis, the coefficient alpha ranges from 0.88 to 0.94, as shown in Table 2. Nunnally and Bernste [14] consider a reliability coefficient of 0.70 or higher for a measuring instrument satisfactory. According to van de Ven and Ferry [15] a reliability coefficient of 0.55 or higher is acceptable, and Osterhof [16] recommends that 0.60 or higher is adequate. Therefore, based on the standards in the literature, the variable items developed for this study are considered reliable.

Convergent validity, according to Campbell and Fiske [17], occurs when the scale items in a given construct move in the same direction and, therefore, correlate strongly with one another. The principal component analysis, which provides a measure of convergent analysis [18], is performed, as reported in Table 2. We have used the Eigen Value as a reference point for observing the construct validity using principal component analysis [19]. Specifically, we have used the Eigen Value One Criterion, also known as the Kaiser Criterion, which means that any component having an Eigen Value greater than one is to be retained ([20] & [21]). In our study, Eigen Value analysis reveals that both the variables completely form a single factor. Therefore, the convergent validity of the variables is sufficient.

We have used minitab-16 to compute coefficient alpha and PCA Eigen values.

**Discussion of the Results**

Traditionally, OSS was designed for technically adept users, thus resulting in a lack of distinction between developers and users. However, OSS is no longer used solely by computer developers; the number of non-technical and novice computer users are growing at a fast pace, highlighting the necessity of understanding and addressing their requirements and expectations [22]. In their empirical study, Raza et al. [23] identify different factors that may be considered by the OSS development community to address usability issues of their projects.

Due to the growing prevalence of novice users, issues relating to usability need to be given research priority. This research examines the way user feedback is handled in OSS development. In response to our first question, thirty percent chose functionality, nineteen percent opted for reliability, thirty percent selected usability as their top priority, seven percent voted for efficiency, eleven percent for maintainability, and the rest three percent picked portability, as shown in Figure 2.

Forty-two percent respondents of our survey stated that they collected user feedback for their project in some form. Seventy-two percent of them affirmed that they made modifications to their project as a result of the collected user feedback.

In software development, the role of usability experts cannot be understated, especially in application software, where end users are the direct audiences. In proprietary software development, large organizations hire experts to share their opinion for making software more usable and acceptable to end users. However, because work in OSS is voluntary and there are fewer resources in OSS development, there are not many usability experts active in the OSS field. However in our survey, seventy-seven percent of respondents (OSS programmers) admitted that they did not consult professional usability experts to improve their project. Out of the twenty-three percent who took opinion of usability experts in some form, only one-third declared that they did make some modifications to their project based on the experts' advice.

*Limitations of the Study & Threats to External Validity*

There are several empirical methods for investigating both software engineering processes and products, including surveys, experiments, metrics, case studies, and field studies [24]. All of these empirical investigations are subject to certain limitations, which is the case with this study.

Generalization of experimental results gets limited because of threats to external validity [25]. In this study, we have taken specific measures to support external validity; for example, we used a random sampling technique to pick the respondents from the population. In addition, we retrieved the data from recognized OSS reporting website, sourceforge.net, which contains a considerable number of projects.

Ethical concerns have also been raised due to the increasing popularity of empirical methodologies in software engineering ([26] & [27]). However, in this study, we have followed the recommended ethical principles to ensure that the empirical analysis would not violate any form of the recommended experimental ethics.

Another limitation of this study is its relatively small sample size. Although we sent our survey to a considerable number of OSS developers in 19 different projects of software, we received only 72 responses. Although the proposed approach has some potential to threaten external validity, we have followed the appropriate research procedures by conducting and reporting tests to improve the reliability and validity of the study.

**Conclusion**

We believe that to achieve users' satisfaction, OSS designers and developers need to understand their expectations and requirements. According to our survey results, seventy percent of OSS developers do not consider usability as their top priority. Similarly, the majority of the respondents declared that they neither collected user feedback nor did they consult usability experts to improve their software. We thus conclude that there is a need to take software usability more seriously by OSS managers and developers. Users' feedback and usability experts' opinions can definitely be used to improve OSS usability.

**References**

[1]. International Standard ISO/IEC 9126-1, 2001, Software Engineering – Product Quality – Part 1: Quality model, 1st edition, pp. 9-10.

[2]. E.S. Raymond, 1999, *The Cathedral and the Bazaar*, O'Reilly, Sebastopol, CA.

[3]. L. Zhao, F. P. Deek and J. A. McHugh, 2010, "Exploratory inspection—a user-based learning method for improving open source software usability," Journal of Software Maintenance and Evolution: Research and Practice, volume 22, number 8, pp. 653–675

[4]. G. Çetin, G. and M. Göktürk, 2008, "A measurement based framework for assessment of usability-centricness of open source software projects," Proceedings of IEEE International Conference on Signal Image Technology and Internet Based Systems, SITIS '08

[5]. T. Otte, R. Moreton, and H.D. Knoell, 2008, "Applied quality assurance methods under the open source development model,"Proceedings of IEEE 32nd International Computer Software and Applications Conference (COMPSAC), pp. 1247-52

[6]. M. Bodker, L. Nielsen, and R.N. Orngreen, 2007, "Enabling user centered design processes in open source communities, Usability and Internationalization, HCI and Culture," Proceedings of 2nd International Conference on Usability and Internationalization, UI-HCII

[7]. N. Bevan, 2009, "International standards for usability should be more widely used," Journal of Usability Studies, volume 4, number 3, pp. 106-113

[8]. D.M. Nichols and M.B. Twidale, 2005, "The usability of open source software," First Monday, volume 8, number 1, http://firstmonday.org/issues/issue8 1/nichols/

[9]. N. Iivari, H. Hedberg and T. Kirves, 2008, "Usability in company open source software context - Initial findings from an empirical case study", Open Source Development, Communities and Quality in IFIP International Federation for Information Processing, volume 275, pp. 359–365

[10]. G. Çetin, D. Verzulli and S. Frings, 2007, "An analysis of involvement of HCI experts in distributed software development: practical issues - Online communities and social computing," Proceedings of 2nd International Conference, OCSC 2007, pp. 32-40

[11]. S.Y.T. Lee, H.W. Kim and S. Gupta, 2009, "Measuring open source software success," Omega, volume 37, number 2, pp. 426-438

[12]. A. Raza and L.F. Capretz, 2010, "Contributors' preference in open source software usability: An empirical study," International Journal of Software Engineering & Applications (IJSEA), volume 1, number 2, pp. 45 – 64

[13]. L.J. Cronbach, 1951, "Coefficient alpha and the internal consistency of tests," Psychometrica, volume 16, pp. 297–334

[14]. J.C. Nunnally and I.A. Bernste, 1994, *Psychometric Theory*, 3rd ed. McGraw Hill, New York

[15]. A.H. van de Ven and D.L. Ferry, 1980, *Measuring and Assessing Organizations*, John Wiley & Son, New York

[16]. A. Osterhof, 2001, *Classroom Applications of Educational Measurement*, Prentice Hall, NJ

[17]. D.T. Campbell and D.W. Fiske, 1959, "Convergent and discriminant validation by the multi-trait multi-method matrix," Psychological Bulletin, volume 56, number 2, pp. 81–105

[18]. A.L. Comrey and H.B. Lee, 1992, *First Course on Factor Analysis*, 2nd Edition Hillsdale

[19]. H.F. Kaiser, 1970, "A second generation little jiffy," Psychometrika, volume 35, pp. 401–417

[20]. H.F. Kaiser, 1960, "The application of electronic computers to factor analysis," Educational and Psychological Measurement, volume 20, pp. 141–151

[21]. J. Stevens, 1986, *Applied Multivariate Statistics for the Social Sciences*, Hillsdale, NJ.

[22]. N. Iivari, 2009, "Empowering the users? A critical textual analysis of the role of users in open source software development,", AI Society., volume 23, number 4, pp. 511-528

[23]. A. Raza, A., L.F. Capretz and F. Ahmed, 2011, "Users' perception of open source usability: An empirical study," Engineering with Computers, http://www.springerlink.com/content/lh738r6k875g574l/

[24]. J. Singer and N.G. Vinson, 2002, "Ethical issues in empirical studies of software engineering," IEEE Transactions on Software Engineering, volume 28, number 12, pp. 1171-1180

[25]. C. Wohlin, P. Runeson, M. Host, M.C. Ohlsson, B. Regnell and A. Wesslen, 2000, *Experimentation in Software Engineering*, Kluwer Academic Publishers, Norwell, MA.

[26]. R.R. Faden, T.L. Beauchamp and N.M.P. King, 1986, *A History and Theory of Informed Consent*, Oxford University Press

[27]. J. Katz, 1972, *Experimentation with Human Beings*, New York: Russell Sage Foundation

Table 1: Survey Questionnaires

| No. | Statement |
|---|---|
| Q-1. | Which quality attribute do you personally consider most important in the software your project develops? Functionality, Reliability, Usability, Efficiency, Maintainability, Portability |
| Q-2. | Does your project collect user feedback? For Example using: |

[] special "support" email address

[] publicly accessible mailing list

[]online forums

[] publicly accessible bug-tracker

[] publicly accessible chat channel

[] other public way of mass-communication

[] online meetings with users

[] in-person meetings with users

[] automatic collection of usage data (e.g. click-through patterns, shortcut-usage)

Q-3. If you answered "yes" to question 2, were any modifications made to your project as a result of the collected user feedback?

Q-4. Do you consult professional usability experts to improve your project?

Q-5. If you answered "yes" to question 4, have any modifications been made to your project as a result of the expert advice?

Table 2: Coefficient Alpha and Principal Component Analysis

| Usability Factors | Item no. | Coefficient α | PCA Eigen value |
|---|---|---|---|
| Users Feedback | 1 – 2 | 0.88 | 1.79 |
| Usability Experts Opinion | 3 – 4 | 0.94 | 1.89 |

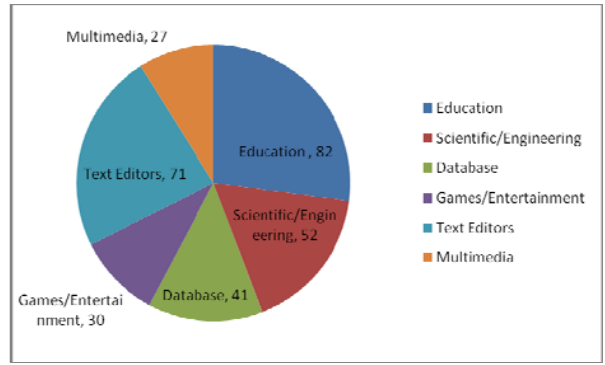

Figure 1 Pie Chart of Software Category

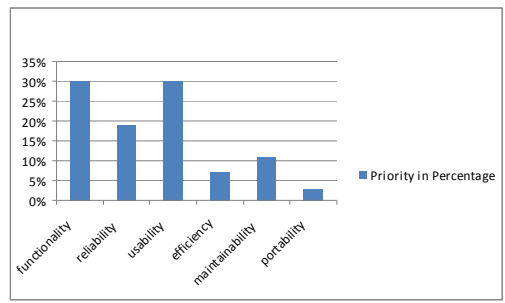

Figure 2 OSS Developers' Priority

## About the authors

**Arif Raza** is a Post-Doctoral Research Fellow, Department of Electrical & Computer Engineering, University of Western Ontario, London, Ontario, Canada. E-mail: araza7@uwo.ca

**Luiz Fernando Capretz** is Professor and Assistant Dean (IT and e-Education), Department of Electrical & Computer Engineering, University of Western Ontario, London, Ontario, Canada. E-mail: lcapretz@uwo.ca